\RequirePackage{lineno} 
\documentclass[aps,prl,twocolumn,superscriptaddress]{revtex4}
\usepackage{amsmath}
\usepackage{amssymb}
\usepackage{graphicx}
\usepackage{color}
\usepackage{float} 

\pdfoutput=1

\newcommand{\eq}{\begin{equation}}
\newcommand{\eeq}{\end{equation}}
\newcommand{\ket}[1]{\left|#1\right\rangle}

\begin{document}

\title{Observation of a Many-Body Dynamical Phase Transition \\ with a 53-Qubit Quantum Simulator}

\author{J. Zhang}
\affiliation{Joint Quantum Institute and Joint Center for Quantum Information and Computer Science, University of Maryland Department of Physics and National Institute of Standards and Technology, College Park, MD 20742}
\author{G. Pagano}
\affiliation{Joint Quantum Institute and Joint Center for Quantum Information and Computer Science, University of Maryland Department of Physics and National Institute of Standards and Technology, College Park, MD 20742}
\author{P. W. Hess}
\affiliation{Joint Quantum Institute and Joint Center for Quantum Information and Computer Science, University of Maryland Department of Physics and National Institute of Standards and Technology, College Park, MD 20742}
\author{A. Kyprianidis}
\affiliation{Joint Quantum Institute and Joint Center for Quantum Information and Computer Science, University of Maryland Department of Physics and National Institute of Standards and Technology, College Park, MD 20742}
\author{P. Becker}
\affiliation{Joint Quantum Institute and Joint Center for Quantum Information and Computer Science, University of Maryland Department of Physics and National Institute of Standards and Technology, College Park, MD 20742}
\author{H. Kaplan}
\affiliation{Joint Quantum Institute and Joint Center for Quantum Information and Computer Science, University of Maryland Department of Physics and National Institute of Standards and Technology, College Park, MD 20742}
\author{A. V. Gorshkov}
\affiliation{Joint Quantum Institute and Joint Center for Quantum Information and Computer Science, University of Maryland Department of Physics and National Institute of Standards and Technology, College Park, MD 20742}
\author{Z.-X. Gong}
\affiliation{Joint Quantum Institute and Joint Center for Quantum Information and Computer Science, University of Maryland Department of Physics and National Institute of Standards and Technology, College Park, MD 20742}
\author{C. Monroe}
\affiliation{Joint Quantum Institute and Joint Center for Quantum Information and Computer Science, University of Maryland Department of Physics and National Institute of Standards and Technology, College Park, MD 20742}
\affiliation{IonQ, Inc., College Park, MD  20740}

\date{\today}

\begin{abstract}
A quantum simulator is a restricted class of quantum computer that controls the interactions between quantum bits in a way that can be mapped to certain difficult quantum many-body problems.  As more control is exerted over larger numbers of qubits, the simulator can tackle a wider range of problems, with the ultimate limit being a universal quantum computer that can solve general classes of hard problems. We use a quantum simulator composed of up to 53 qubits to study a non-equilibrium phase transition in the transverse field Ising model of magnetism, in a regime where conventional statistical mechanics does not apply.  The qubits are represented by trapped ion spins that can be prepared in a variety of initial pure states. We apply a global long-range Ising interaction with controllable strength and range, and measure each individual qubit with near 99\% efficiency.  This allows the single-shot measurement of arbitrary many-body correlations for the direct probing of the dynamical phase transition and the uncovering of computationally intractable features that rely on the long-range interactions and high connectivity between the qubits.
\end{abstract}
\maketitle

There have been many recent demonstrations of quantum simulators with varying numbers of qubits and degrees of individual qubit control~\cite{trabesinger2012nature}. For instance, small numbers of qubits stored in trapped atomic ions~\cite{zhang2017, Jurcevic2016} and superconducting circuits~\cite{barends2015digital} have been used to simulate various magnetic spin or Hubbard models with individual qubit state preparation and measurement. Large numbers of atoms have simulated similar models, but with global control and measurements~\cite{garttner2017measuring} or with correlations that only appear over a few atom sites~\cite{kuhr2016quantum}. An outstanding challenge is to increase qubit number while maintaining individual qubit control and measurement, with the goal of performing simulations or algorithms that cannot be efficiently solved classically.  Atomic systems are excellent candidates for this scaling, because their qubits can be made virtually identical, with flexible and reconfigurable control through external optical fields and high initialization and detection efficiency for individual qubits.  Recent work with neutral atoms~\cite{Labuhn2016,bernien2017probing} has demonstrated many-body quantum dynamics with up to 51 atoms coupled through van der Waals Rydberg interactions, and the current work presents the optical control and measurement of a similar number of atomic ions interacting through their long-range Coulomb-coupled motion.

We perform a quantum simulation of a dynamical phase transition (DPT) with up to 53 trapped ion qubits.  The understanding of such nonequilibrium behavior is of great interest to a wide range of subjects, from social science~\cite{reia2016effect} and cellular biology \cite{davies2011cancer} to astrophysics~\cite{Zurek1996} and quantum condensed matter physics~\cite{hinrichsen2006non}. Recent theoretical studies of DPT~\cite{Yuzbashyan2006, Sciolla2010, Calabrese2011, Mitra2012, Heyl2013, Heyl2014, Heyl2017, Zunkovic2016} involve the transverse field Ising model (TFIM), the quintessential model of quantum phase transitions~\cite{SachdevQPTBook}.  A recent experiment investigated a DPT with up to 10 trapped ion qubits, where the transverse field dominated the interactions~\cite{Jurcevic2016}. These studies have considered long-time spin relaxation dynamics~\cite{Yuzbashyan2006, Sciolla2010, Calabrese2011, Halimeh2017, Zunkovic2016} and non-analytic time evolution of non-local quantities~\cite{Heyl2013, Heyl2014, Heyl2017, Zunkovic2016, Jurcevic2016}. 

\renewcommand{\thefigure}{1}
\begin{figure*}[t]
	\includegraphics[width=1.8\columnwidth]{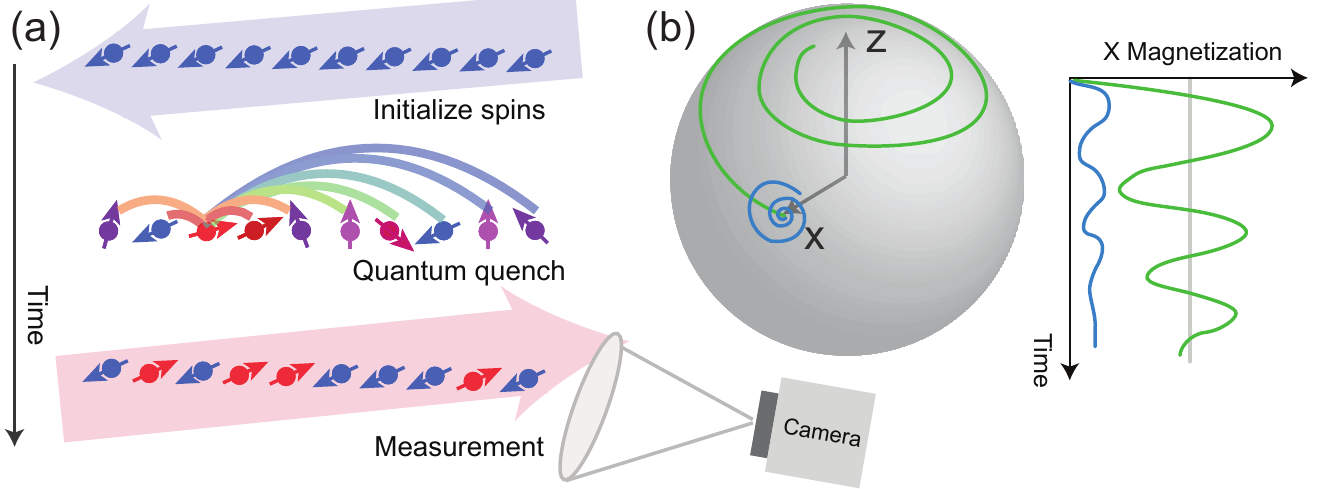}
	\caption{{\bf Illustration of the DPT from a quantum quench}. We subject a system of interacting spins to a sudden change of the Hamiltonian and study the resulting quantum dynamics. (a) An isolated spin system is prepared in a product state, and an Ising spin-spin interaction is suddenly turned on, along with a tunable transverse magnetic field (see text for details). At the end of the evolution, we measure the spin magnetizations along the initial spin orientation direction. (b) A Bloch-sphere representation~\cite{Zunkovic2016} of the average spin magnetization. Spins are initially fully polarized along the longitudinal $x$ direction of the Bloch sphere, and evolve with Ising interactions along $x$ competing with the transverse field along $z$, resulting in oscillations and relaxations. Blue curves illustrate the quench dynamics with a low transverse field; green curves indicate the dynamics with a large transverse field across criticality.}
\label{fig:concept}
\end{figure*}

In this experiment, we employ a quantum quench--a sudden change in the system Hamiltonian--to bring a collection of interacting trapped ion qubits out of equilibrium~\cite{Richerme2014, Jurcevic2014, Neyenhuis2016, Jurcevic2016}. The theoretical description of the dynamics is made difficult by the population of exponentially many excited states of the many-body spectrum, typically accompanied by massive entanglement between the qubits.  Given the long-range interactions between the qubits, the entanglement growth is generally much faster~\cite{Gong2017} than in locally connected systems~\cite{Labuhn2016, bernien2017probing}, making the classical simulation of the quench dynamics even more challenging.  The nature of the long-range Ising interaction also leads to unique dynamical features and an emergent higher dimensionality of the system~\cite{Dyson1969, Zunkovic2016, Maghrebi2017}.

\renewcommand{\thefigure}{2}
\begin{figure*}[ht!]
\includegraphics[width=2.05\columnwidth]{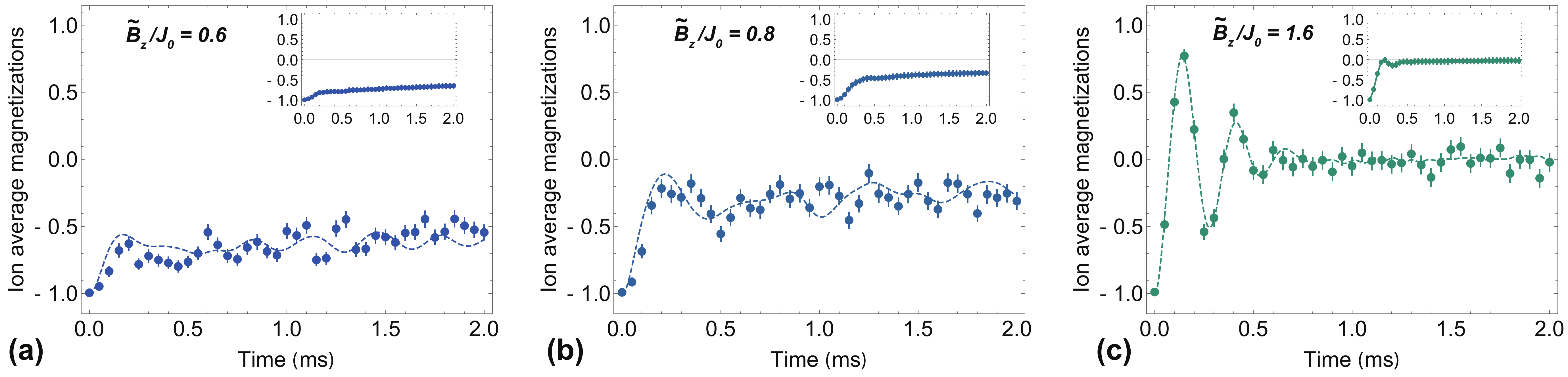}
\caption{\textbf{Real-time spin dynamics after a quantum quench of 16 spins in an Ising chain.} (a) Polarized spins evolve under the long-range Ising Hamiltonian with a small transverse field ($\tilde{B_z}/J_0 = 0.6$). The broken symmetry given by the initial polarized state is preserved during the evolution. (b) When the transverse field is increased ($\tilde{B_z}/J_0 = 0.8$), the dynamics shows a faster initial relaxation, before settling to a non-zero plateau. (c) Under larger transverse fields ($\tilde{B_z}/J_0 = 1.6$), the Larmor precession takes over, and the spins oscillate and relax to zero average magnetization. The dashed lines are numerical simulations based on exact diagonalization. Insets: cumulative time-averages of the spin magnetization, smoothing out temporal fluctuations and showing the plateaus. Each point is the average of 200 experimental repetitions. Error bars are statistical, computed from quantum projection noise and detection infidelities as described in Appendix B.}
\label{fig:SpinOscillations}
\end{figure*}
\renewcommand{\thefigure}{3}
\begin{figure}[h!]
\includegraphics[scale=0.4]{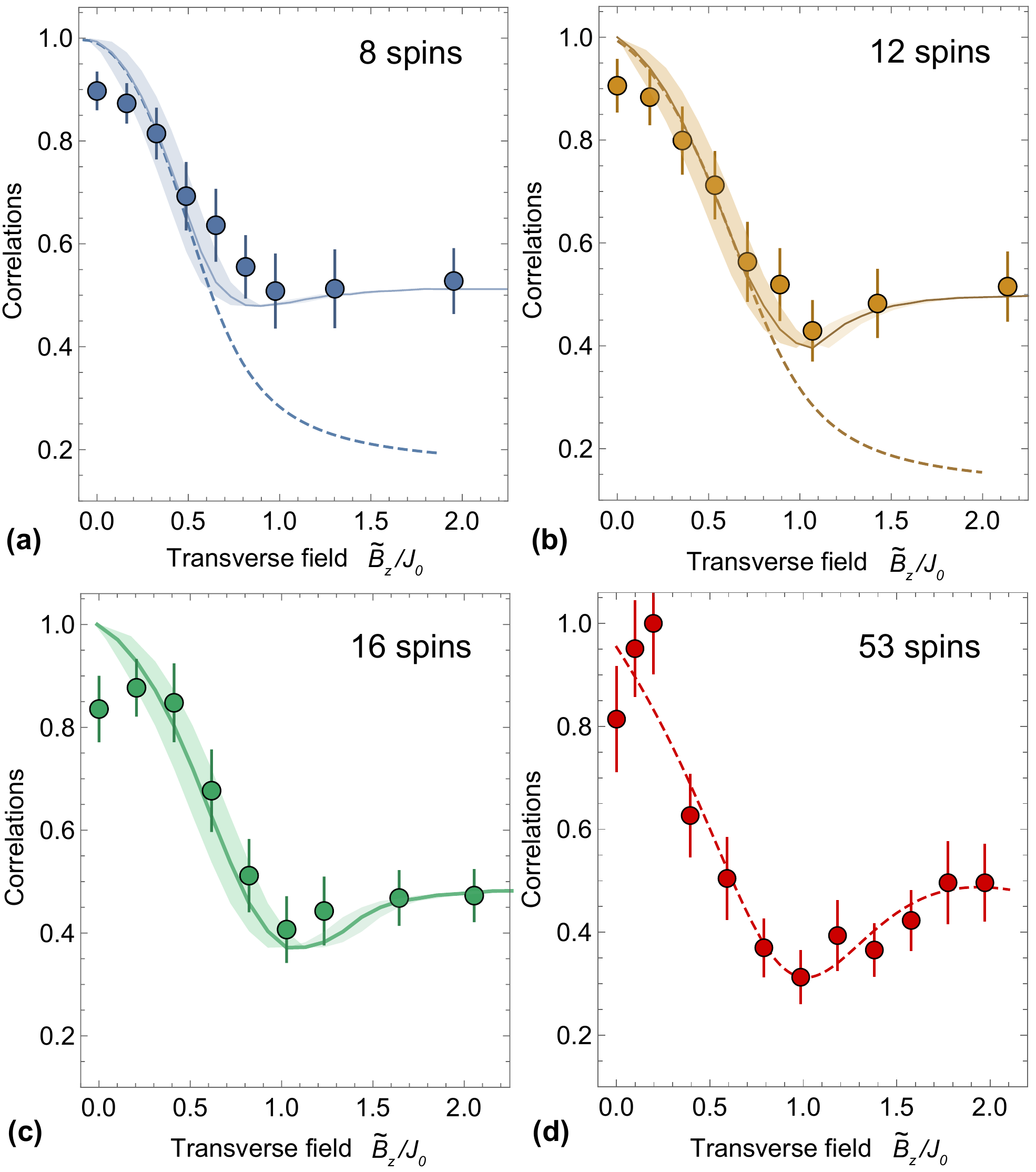}
\caption{\textbf{Two-body Correlations.} Long-time averaged values of the two-body correlations $C_2$ over all pairs of spins, for different numbers of spins in the chain. Statistical error bars are $\pm$ one standard deviation from measurements covering 21 different time steps. Solid lines in (a)-(c) are exact numerical solutions to the Schr\"odinger equation, and the shaded regions take into account uncertainties from experimental Stark shift calibration errors. Dashed lines in (a) and (b) are calculations using a canonical (thermal) ensemble with an effective temperature corresponding to the initial energy density. For N=53 spins in (d), the correlations are uniformly degraded from residual Stark shifts across the ion chain, so in this case we normalize to the maximum correlation at small field (see Appendix D).
Exact diagonalization for N=53 spins is out of reach, so we instead fit the experimental data to a Lorentzian function with linear background, shown by the dashed line.}
\label{fig:Correlations}
\end{figure}
\renewcommand{\thefigure}{4}
\begin{figure*}[ht!]
	\includegraphics[width=2.05\columnwidth]{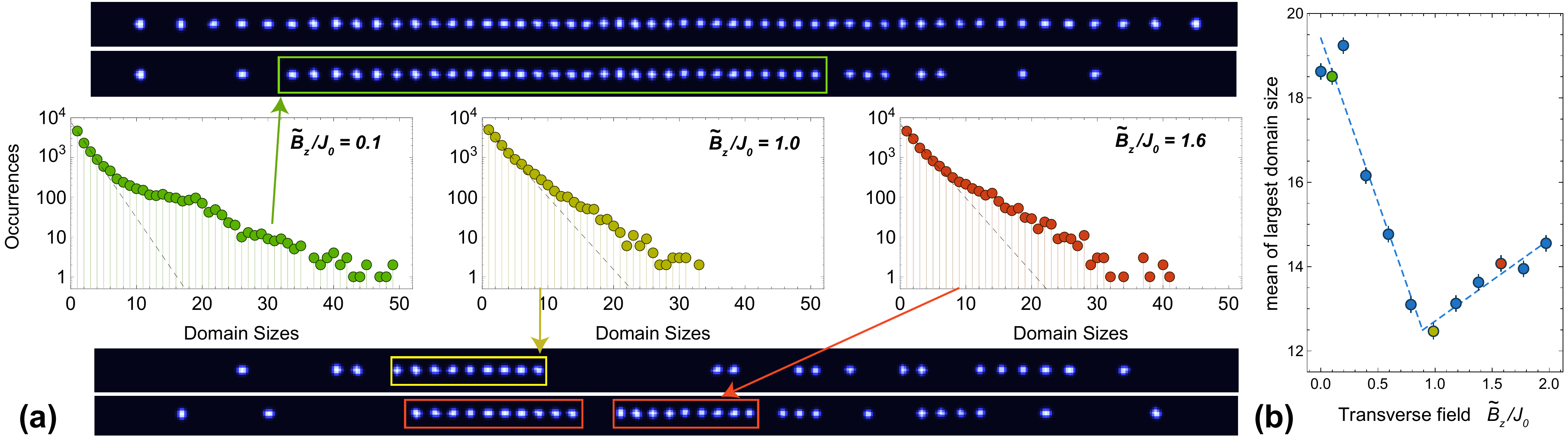}
	\caption{{\bf Domain statistics and reconstructed single shot images of 53 spins}. (a) Top and bottom: reconstructed images based on binary detection of spin state (see Appendix E).The top image shows a chain of 53 ions in bright spin states. The other three images show 53 ions in combinations of bright and dark spin states. Center: statistics of the sizes of domains, or blocks with spins pointing along the same direction. Histograms are plotted on a logarithmic scale, to visualize the rare regions with large domains. Dashed lines are fits to exponential functions, which could be expected for infinite-temperature thermal state. Long tails of deviations are clearly visible, and varies depending on $\tilde{B_z}/J_0$. (b) Mean of the largest domain sizes in each single experimental shot. Error bars are the standard deviation of the mean (see Appendix F). Dashed lines represent a piecewise linear fit, from which we extract the transition point (see text). The green, yellow, and red data points correspond to the transverse fields shown in the domain statistics data on the left.} 
\label{fig:longChains}
\end{figure*}

We experimentally implement a quantum many-body Hamiltonian with long-range Ising interactions and flexible tuning parameters~\cite{Porras2004, Kim2009}.  As outlined in Fig.~\ref{fig:concept}, we initialize the qubits (effective spin-1/2 systems) in a product state all polarized along the $x$ direction of the Bloch sphere, and suddenly turn on the TFIM Hamiltonian given by ($h=1$)
\begin{equation} 
H = \sum_{i<j} J_{ij} \sigma^x_i \sigma^x_{j}+B_z\sum_i\sigma_i^z.
\label{eq:model}
\end{equation}
Here $\sigma_i^\gamma$ ($\gamma=x,y,z$) is the Pauli matrix acting on the $i^\text{th}$ spin along the $\gamma$ direction of the Bloch sphere, $J_{ij}$ is the Ising coupling between spins $i$ and $j$, and $B_z$ denotes the transverse magnetic field, which acts as the control parameter for crossing dynamical criticality in the DPT.

The right panel of Fig.~\ref{fig:concept} shows a simplified Bloch-sphere representation of the DPT dynamics. The spins quickly evolve from the longitudinally polarized initial state, and then either precess about a large transverse magnetic field (green curves in Fig.~\ref{fig:concept}), or stay pinned near the initial conditions when the transverse field is small (blue curves in Fig.~\ref{fig:concept}). 

To implement the quantum Hamiltonian (see Appendices B-C), each spin in the chain is encoded in the $^2$S$_{1/2}\ket{F=0,m_F=0} \equiv \ket{\downarrow}_z$ and $\ket{F=1,m_F=0} \equiv \ket{\uparrow}_z$ hyperfine ``clock" states of a $^{171}$Yb$^+$ ion and separated by a frequency of $\nu_0 = 12.642821$ GHz.  We store a chain of up to $N=53$ ions in a linear rf Paul trap, as described in Appendix A~\cite{Kim2009} and initialize the qubits in the product state $\left | \downarrow \downarrow \cdots \downarrow \right \rangle_x$,
where $\ket{\downarrow}_x \equiv \ket{\downarrow}_z+\ket{\uparrow}_z$. Spin-spin interactions are generated by spin-dependent optical dipole forces from an applied laser field, which give rise to tunable long-range Ising couplings that fall off approximately algebraically as $J_{ij} \approx J_\textrm{0}/|i-j|^\alpha$~\cite{Porras2004, Kim2009, Islam2013}. The power-law exponent $\alpha$ is set between  $0.8-1.0$ in the experiment, and the maximum interaction strengths are $J_0 = $(0.82, 0.56, 0.38, 0.65) kHz, for (8, 12, 16, and 53) spins, respectively. The transverse field is generated by a controllable Stark shift of the spin qubit splitting from the same laser field, as described in Appendix C.

We finally measure the magnetization of each spin $\langle \sigma_i^x \rangle$.  We rotate all the spins by $\pi/2$ about the $y$-axis of the Bloch sphere (exchanging $\sigma_i^x \leftrightarrow \sigma_i^z$) and then illuminate the ions with resonant radiation and collect the $\sigma_i^z$-dependent fluorescence on a camera with site-resolved imaging~\cite{Islam2013}. We estimate a spin detection efficiency of $\sim 99\%$ for each qubit (see Appendix E), providing access to all possible many-body correlators in a single shot. 

The simplest observable of quench dynamics, after evolving the system under the TFIM for time $t$, is the average magnetization of the spins along $x$,
$\langle \sigma^x(t) \rangle = \sum_i \langle \sigma_i^x(t) \rangle/N$.
Figure~\ref{fig:SpinOscillations} shows the measured average magnetization for $N=16$ spins through $2\pi J_0 t = 4.8$, for different values of the transverse field.  We formulate a renormalized field $\tilde{B}_z$, to account for the divergence of the energy density of the long-range Ising interactions, so that the ratio $\tilde{B}_z/J_o$ is meaningful in the thermodynamic limit (see Appendix C and Ref. ~\cite{Kac1969}).  This allows a fair comparison of the DPT for different numbers of spins in the chain.

The evolution of the time-dependent magnetization separates into two distinctive regimes: one that breaks the $\mathbb{Z}_2$ symmetry 
$(\sigma^{x,y}_i \rightarrow -\sigma^{x,y}_i)$
of the Ising Hamiltonian (Fig.~\ref{fig:SpinOscillations}a), as was explicitly set by the initial conditions; and one that restores this symmetry (Fig.~\ref{fig:SpinOscillations}c), where the intermediate time dynamics oscillates around and relaxes to zero average magnetization. Between these two regimes we observe a relaxation to a non-zero steady value (Fig.~\ref{fig:SpinOscillations}b). Cumulative time-averages $\overline{\langle \sigma^x \rangle}(t)= \int_0^t \langle \sigma^x(\tau) \rangle d\tau/t $ (insets in Fig.~\ref{fig:SpinOscillations}) clearly reveal the long-time magnetization plateaus.

The DPT is expected to occur between the small and large transverse field regimes, where the spin alignment changes abruptly from ferromagnetic to paramagnetic in the long time limit as shown in Fig. \ref{fig:concept}. This phase transition is well-established for $\alpha=0$, as shown in Appendix G. Strong numerical evidence shows that such a transition will survive \cite{Zunkovic2016,Halimeh2017} for the small values of $\alpha$ chosen in our experiments, but not for $\alpha=\infty$ where interactions are nearest-neighbor only.

Further signatures of the DPT are observed by measuring the spatially averaged two-spin correlations $C_2=\sum_{i,j} \langle \sigma_i^x \sigma_j^x\rangle/N^2$. From the behavior of the magnetizations described above, we expect that $C_2 \rightarrow 1$ for small $\tilde{B}_z$ and $C_2\rightarrow 1/2$ for large $\tilde{B}_z$ at long times, since the collective spin precesses around the z axis and $C_2$ oscillates between one and zero. Figure~\ref{fig:Correlations} shows the cumulative time-averaged correlations. Near the critical value of $\tilde{B}_z$, we observe the emergence of a dip in $C_2$ (Fig. \ref{fig:Correlations}, which is a direct signature of the DPT.  The sharpening of the dip for larger system sizes is not strong, which may be due to a logarithmic finite-size scaling (see Appendix G). 

For a non-integrable system such as the long-range TFIM studied here, it might be conjectured that the spins eventually reach a thermal distribution~\cite{Rigol2012}. However, we find that this is only true for small $\tilde{B}_z$ (Fig. \ref{fig:Correlations}a-b). We note that the thermal values of the correlator $C_2$ do not exhibit a dip or show signatures of a phase transition with varying $\tilde{B}_z/J_\textrm{0}$ for system sizes that we are able to model numerically. Interestingly, thermalization appears to break down in this quenched system, which we suspect is a consequence of the inherent long-range nature of the Ising interactions \cite{Campa2008}. 

We further explore many-body dynamical properties of this system by investigating higher-order correlations, which are even harder to calculate classically~\cite{DMRG}. Through high-efficiency single-shot state detection of all of the spins, we directly measure higher-order correlation observables. Single-shot images for $N=53$ spins are shown in Fig.~\ref{fig:longChains}a and are reconstructed from binary thresholding and image convolution of the ion chain fluorescence distribution (Appendix E). The analysis of these binary strings gives direct information of correlations up to arbitrary order. 

The occurrence of long domains of correlated spins in the state $\ket{\uparrow}_x$ (fluorescing spins) signifies the fully polarized initial state, where the correlations in the initial state are largely preserved by the interactions. With an increasing transverse field, the absence of spin-ordering is reflected by exponentially small probabilities for observing long strings. We plot the domain length statistics in Fig.~\ref{fig:longChains}a at late times (see Appendix F), for three example transverse field strengths, $\tilde{B}_z/J_0 = $ (0.1, 1.0, 1.6). The dashed lines in Fig.~\ref{fig:longChains}a are fits to exponentials on the histogram of domain sizes. The rare occurrence of especially large domains (e.g. the red boxes in Fig.~\ref{fig:longChains}a) shows the existence of many-body high-order correlations, where the order is given by the length of the domain. We plot the mean of the largest domain size in Fig.~\ref{fig:longChains}b, as a function of the normalized transverse field strength. The average longest domain size ranges from 12 to 20, and shows a sharp transition across the critical point of the DPT. We fit this observable to a piecewise linear function, and extract the critical point to be $\tilde{B}_z/J_0$ = 0.89(7). For more details, see Appendix F.

The DPT studied here, with up to 53 trapped ion qubits, is the largest quantum simulation ever performed with high-efficiency single shot individual qubit measurements.  This gives access to arbitrary many-body correlators that carry information that is difficult or impossible to model classically. This experimental platform can be extended to tackle provably hard quantum problems such as Ising sampling \cite{IsingSampling}.  Given an even higher level of control over the interactions between spins, as already demonstrated for smaller numbers of trapped ion qubits \cite{Debnath2016}, this same system can be upgraded to a universal quantum computer.

\newpage

\section{Appendix A: Confinement of long ion chains}

The ion chain is confined in a  3-layer linear Paul trap with $\nu_{cm} =4.85$~MHz transverse center-of-mass motional frequency \cite{Kim2009}.  The harmonic axial confinement is kept low enough so that the lowest energy conformation of the ions is linear; for $8-16$ ions the axial center-of-mass frequency is $\sim 400$ kHz and for $53$ ions it is $\sim 200$ kHz. The ion spacing is anisotropic across the chain, with typical spacings of $1.5$ $\mu$m at the center of the chain and $3.5$ $\mu$m at either end \cite{James1998}.

The effective lifetime of an ion chain is limited by Langevin collisions with the residual background gas in the UHV apparatus \cite{NIST_bible}, which in general re-orders the crystal but can also melt the crystal and even ultimately eject the ions from rf-heating or other mechanisms.   This can be mitigated by quickly re-cooling the chain, and we expect that occasionally the crystal returns without notice.  Rarely, such collisions with the background gas are inelastic, either populating the $^{171}$Yb$^+$ ion in the metastable $F_{7/2}$ state or forming a YbH$^+$ molecule. The 355 nm Raman laser quickly returns the ions back to their atomic ground state manifold, with a small probability of creating doubly-charged ions.  The mean time between Langevin collisions is expected to be of order 1 collision per hour per trapped ion, and we expect that the mean lifetime for a chain of ions might therefore scale inversely with the number of ions. For 53 ions we observe an average lifetime of about 5 minutes.  However, we observe rare events where a long ion chain survives for about 30 minutes.  We speculate that either the chain is consistently re-captured instantaneously, or the local pressure in the chamber is anomalously low during these periods.  Because we can load an ion chain in under 1 minute, this enables a reasonable duty-cycle of collecting data.  

\section{Appendix B: State preparation}

Two off-resonant laser beams at 355 nm globally address the ions and drive stimulated Raman transitions between the two hyperfine qubit clock states $^2S_{1/2}\ket{F=0, m_F=0}$ and $\ket{F=1,m_F=0}$. The Raman beatnotes are provided by the frequency comb from the mode-locked laser, resulting in coherent qubit rotations~\cite{Olmschenk2007,Hayes2010}. The ion chain is about 100 $\mu m$ in length, and the beams are focused to a 200 $\mu m$ full width half maximum along the ion chain, resulting in a $30\sim40\%$ intensity imbalance between the center and edges of the chain. In order to prepare each individual ion in the $\ket{\downarrow}_x \equiv \ket{\downarrow}_z+\ket{\uparrow}_z$ state, we first optically pump the ions in the $\ket{\downarrow}_z$ state with a $99.9\%$ efficiency \cite {Olmschenk2007} and then we apply a $\pi/2$ rotation around $y$ axis. However, if we use a square pulse, the beam inhomogeneity leads to an imperfect state preparation.

To mitigate intensity imperfections across the chain, we employ a BB1 dynamical decoupling pulse sequence~\cite{Brown2004} (written for each spin $i$):
\begin{equation}
U_1 (\pi/2) =e^{-i\frac{\pi}{2}\sigma_i^{\theta}} e^{-i\pi\sigma_i^{3\theta}} e^{-i\frac{\pi}{2}\sigma_i^{\theta}} e^{-i\frac{\pi}{4}\sigma_i^{y}} , \nonumber \\
\end{equation}where, in addition to the $\pi/2$ rotation $e^{-i\frac{\pi}{4}\sigma_i^{y}}$, three additional rotations are applied: a $\pi$-pulse along an angle $\theta = \textrm{cos}^{-1}(-1/16) = 93.6^{\circ}$, a $2\pi$-pulse along $3\theta$, and another $\pi$-pulse along $\theta$, where the axes of these additional rotations are in the $x$-$y$ plane of the Bloch sphere with the specified angle referenced to the $x$-axis. With this scheme, we measure a state preparation fidelity of up to 99\% for the well-compensated ions, and an average fidelity of 93\%, limited by the ions at the edges of the chain. 

\section{Appendix C: Generating the Ising Hamiltonian}
We generate spin-spin interactions by applying a spin dependent optical dipole force induced by the global Raman beams, which are aligned with a wavevector difference $\Delta k$ along a principal axis of transverse motion \cite{Kim2009}. Two beatnotes of the non-copropagating Raman beams are tuned near the transverse upper and lower motional sideband frequencies at $\nu_0 \pm \mu$, in the usual M\o{}lmer-S\o{}rensen configuration~\cite{MS}. 
In the Lamb-Dicke regime, this gives rise to the Ising-type Hamiltonian \cite{Kim2009}  in Eq. (1) with Ising coupling between ions $i$ and $j$,
\begin{equation}
\label{Jij}
J_{ij} = \Omega^2\nu_{R} \sum_{m}\frac{b_{im}b_{jm}}{\mu^2-\nu_m^2}
\approx \frac{J_0}{|i-j|^\alpha}.
\end{equation}
Here $\Omega$ is the global (carrier) Rabi frequency, $\nu_R= h \Delta  k^2/(8\pi^2M)$ is the recoil frequency, $b_{im}$ is the normal mode transformation matrix of the $i$-th ion with the $m$th normal mode $(\sum_i |b_{im}|^2=\sum_m |b_{im}|^2 =1)$ \cite{James1998}, $M$ is the mass of a single ion, and $\nu_m$ is the frequency of the $m$-th normal mode.  Here, the beatnote frequency detuning $\mu$ is assumed to be sufficiently far from all sidebands, or $|\mu-\nu_m| \gg \Omega b_{im}\sqrt{\nu_R/\nu_m}$, so that the spins only couple through the motion virtually and no phonons are produced.  

The approximate power-law exponent in Eq. \ref{Jij} can be tuned between $0<\alpha<3$ in principle, but in practice we are restricted to $0.5<\alpha<1.8$ in order to avoid motional decoherence and experimental drifts. To keep $\alpha$ roughly constant across the different system sizes we adjust the sideband detuning $\delta_m=\mu-\nu_m$ to the values $\delta_m = \pm (56,69,82,60)$~kHz for $N=(8,12,16,53)$, respectively, and we set the Rabi frequencies so that the respective nearest-neighbor Ising couplings are $J_0 = (0.82, 0.56,0.38, 0.65)$ kHz. In this work we have $\alpha\approx0.8$ for $N = 8-16$, and $\alpha\approx 1$ for $N=53$. 

With $\alpha<1$ the long range interaction term in the Hamiltonian (1) is super-extensive for a 1D linear chain. In order to have a well defined thermodynamic limit of the Hamiltonian, the couplings are typically rescaled to $\tilde{J}_{ij} = J_{ij}/\mathcal{N}$ using the Kac normalization constant~\cite{Kac1969}
\begin{equation}
\label{eq:}
\mathcal{N}=\frac{1}{N}\sum_{i,j}\frac{J_{ij}}{J_0}.\nonumber
\end{equation}
Since all our observables are a function of the ratio of the field to the Ising coupling strength $B_z/J_0$, we instead equivalently renormalize the magnetic field using $\tilde{B_z}=\mathcal{N} B_z $ and retain the original form of the Ising coupling.
\renewcommand{\thefigure}{5}
\begin{figure*}[t]
\includegraphics[width=2\columnwidth]{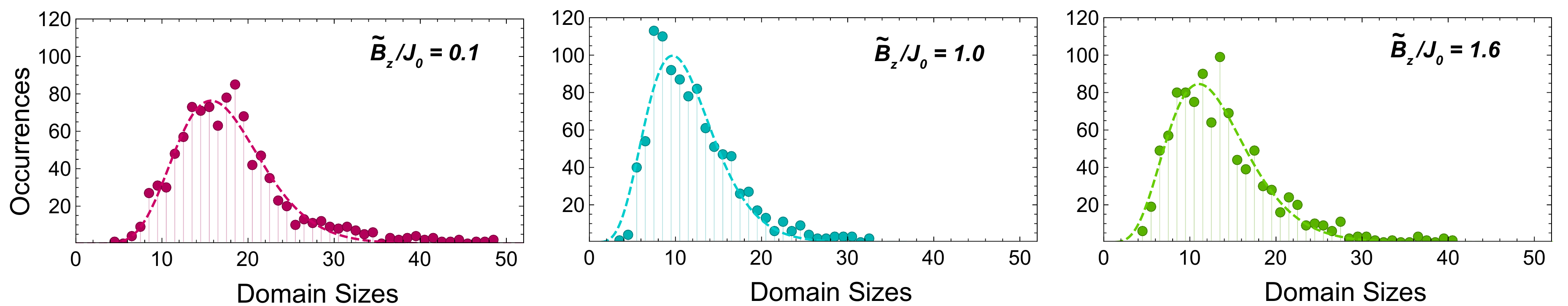}
\caption{{\bf Distributions of the largest domain size.} Statistics of the largest domain size in each experimental shot (200 experiments for each of the last 5 time steps). Considering only the largest domains of each shot eliminates undesirable biasing toward small domain sizes present in Fig.~\ref{fig:longChains}a. Domain sizes are related to many-body correlators, where a domain size of N corresponds to an N-body correlator. Dashed lines are fits to a two parameter Gamma distribution proportional to $e^{-x / \beta} x^{\alpha-1}$, with shape parameter $\alpha$ and scale parameter $\beta$. }
\label{fig:OutlierDist}
\end{figure*}

\section{Appendix D: Generating the transverse magnetic field}

In order to generate the effective magnetic field, we asymmetrically adjust the two Raman beatnotes to  $\nu_0\pm\mu + B_z$  resulting in a uniform effective transverse magnetic field of $B_z$ in Eq. (1) (not yet Kac-renormalized as described above).
 
To induce the quantum quench, the sidebands are switched on in about 100 ns using acousto-optic modulators (AOMs), which control the detuning  and amplitude of the Raman beatnotes. These two beatnotes correspond to different beam angles out of the AOM, so we image these beams onto the ion chain in order to maximize the overlap of all frequency components. We measure a residual effective linear gradient of magnetic field across the chain, resulting from a fourth-order Stark shift gradient \cite{Lee2016} arising from the non-perfect overlap of the two beatnotes. This effect is measured to be $\Delta B_z = \pm 0.65$ kHz end-to-end on a 16 ion chain, and was included in the numerics. This gradient is dominated by uniform magnetic fields $B_z>2$ kHz, but it still plays a role at zero or small magnetic fields causing an effective depolarization of the initial state $\left | \downarrow \downarrow \cdots \downarrow \right \rangle_x$. Additional spin-depolarization errors can be caused by Stark shift fluctuations or residual spin-phonon coupling, and are likely responsible for the slight decay seen in Fig.~\ref{fig:SpinOscillations}a of the main text.

\section{Appendix E: Single shot detection and image processing}

We detect the ion spin state by globally rotating all the spins into the measurement basis (composite (BB1) $\frac{\pi}{2}$ pulse as describe above, to rotate x basis into z basis), followed by the scattering of resonant laser radiation on the $^2$S$_{1/2}\ket{F=1} - ^2$P$_{1/2}\ket{F=0}$ cycling transition (wavelength near 369.5 nm and radiative linewidth $\gamma/2\pi\approx 20$ MHz). The $\ket{\uparrow}_z$ ``bright" state fluoresces strongly while the $\ket{\downarrow}_z$  ``dark" state fluoresces almost no photons because the laser is far from resonance~\cite{Olmschenk2007}. 

The fluorescence of the ion chain is imaged onto an EMCCD camera (Model Andor iXon 897) using an imaging objective with 0.4 numerical aperture and a magnification of 60.  The fluorescence of each ion covers roughly a 5x5 array of pixels on the EMCCD. After collecting the fluorescence for an integration time of 300 $\mu$s, we collect a mean of about 20 photons per bright ion, distributed in a circular region of interest (ROI) around the center of the ion position.  In every single shot, we use a simple binary threshold to determine the state of each ion ($\left | \downarrow \right \rangle_z$ or$\left | \uparrow \right \rangle_z$), providing a binary detection of the quantum state of any ion with near 99$\%$ accuracy.  The residual 1\% errors include off-resonant optical pumping of the ion between states during detection, readout noise and background counts, and crosstalk between adjacent ions.

The individual ion ROI areas on the camera are determined from periodically acquiring diagnostic images, where a resonant repumper laser is applied to cause each ion to fluoresce strongly regardless of its state.  The signal to background noise ratio in the diagnostic shots is larger than 100, yielding precise knowledge of the center locations. Ion separations range from 1.5 um to 3.5 um depending on the trap settings and the distance from the chain center, and are always much larger than the resolution limit of the diffraction-limited imaging system (500$^{+100}_{-0}$ nm Airy ring radius projected at the ion position). We utilize the pre-determined ion centers to process the individual detection shots and optimize the integration area on the EMCCD camera to collect each ion's fluorescence while minimizing crosstalk.  We estimate crosstalk to be dominated by nearest-neighbor fluorescence, which can bias a dark ion to be erroneously read as bright with less than $1\%$ probability. 

\section{Appendix F: Domain size statistical data analysis}
Here we present the detailed analysis of the domain statistics presented in Fig.~\ref{fig:longChains}. The raw domain statistics are analyzed from the binary tally of bright and dark ions, and sorting them into domains with consecutive spins up (bright) or down (dark). The collection of all 200 experimental repetitions for the last 5 time steps (out of 21 time steps in total) are treated equally, and results into the statistics given in Fig.~\ref{fig:longChains}a. 

To analyze the large domains, or the outliers of the distributions in Fig.~\ref{fig:longChains}, we find the largest domain in each single shot, and plot the statistical distribution in  Fig. 5. In the main text the mean (standard error of the mean) are used to extract the data (error bars) presented in Fig.~\ref{fig:longChains}b. This has an underlying assumption that the central-limit theorem holds for our largest domain size statistics. 
In addition, we analyze the distribution in the actual data, and fit the histogram to a two parameter Gamma distribution, shown as the dashed lines in Fig. 5. From the fit parameters we can extract the mean, taking the skewness of the distribution into account. This systematically shifts the largest domain size by about 1 for all the datasets, and a piecewise fit similar to that described in the main text yields the critical point $\tilde{B_z}/J_0 =$ 0.92(7) from this alternative data analysis method, in good agreement with that obtained in the main text.

\section{Appendix G: Analytical study of the DPT for $\alpha=0$}
In this section, we show analytically that in the limit where $\alpha\rightarrow0$ ($J_{ij}=J_{0}$ for $i\ne j$), the spatially averaged two-point correlation $C_2=\sum_{i,j}\langle\sigma_{i}^{x}\sigma_{j}^{x}\rangle/N^2$ measured in the experiment will undergo a DPT when $\tilde{B}_{z}/J_{0}$ crosses unity, in the thermodynamic and long time limit. The case $\alpha\sim 1$ in our experiment cannot be treated analytically or numerically for large system sizes, but appears to have qualitatively similar dynamics with the $\alpha=0$ case treated here analytically. 

We first rewrite the Hamiltonian for $\alpha=0$ using collective spin operators $\Sigma^{x,y,z}=\sum_{i=1}^{N}\sigma_{i}^{x,y,z}$,
\begin{equation}
H_{0}=\frac{J_{0}}{N}(\Sigma^{x})^{2}+\tilde{B}_{z}\Sigma^{z}.
\label{eq:s1}
\end{equation}
We then normalize the Ising interactions to make $H_{0}$ extensive (See Apprendix C), which allows a well-defined thermodynamic limit. According to the Heisenberg equation, we have (setting $h=1$)
\begin{equation}
\frac{d\sigma^{x}}{dt}=i[H_{0},\Sigma^{x}]=-2\tilde{B}_{z}\Sigma^{z}.
\end{equation}
We note that the thermodynamic ($N\rightarrow\infty$) limit coincides with the semiclassical limit for the Hamiltonian in Eq.\,(\ref{eq:s1}). Thus we can assign to the values of $\Sigma^{x,y,z}$ classical vectors of length $N$ on a Bloch sphere, i.e. $(\Sigma^{x},\Sigma^{y},\Sigma^{z})=N(\cos\theta,\sin\theta\sin\phi,\sin\theta\cos\phi)$. The above equation of motion can then be reduced to:
\begin{equation}
\frac{d\theta}{dt}=2\tilde{B}_{z}\sin\phi,
\end{equation}
together with the equation $\cos\phi=(J_{0}/\tilde{B}_{z})\sin\theta$ that comes from energy conservation. 

Given the initial state $\theta(t=0)=0$, the dynamics of the correlation $C_2=\cos^{2}\theta$ can be obtained analytically. In the long time limit, we find that the time-averaged value of $\overline{C_2}(\infty)\equiv\lim_{T\rightarrow\infty}\frac{1}{T}\int_{0}^{T}\langle C_2(t)\rangle dt$ is
\begin{equation}
\overline{C_2}(\infty)\equiv\frac{\int_{0}^{\xi}\frac{\cos^{2}\theta d\theta}{2\sqrt{(\tilde{B}_{z}/J_{0})^{2}-\sin^{2}\theta}}}{\int_{0}^{\xi}\frac{d\theta}{2\sqrt{(\tilde{B}_{z}/J_{0})^{2}-\sin^{2}\theta}}}\label{eq:C1} ,
\end{equation}
where $\xi=\sin^{-1}[\min(|\tilde{B}_{z}/J_{0}|,1)]$. We have plotted $\overline{C_2}(\infty)$ as a function of $\tilde{B}_{z}/J_{0}$ in Figure 6. A sharp dip is observed, confirming the existence of the dynamical phase transition.

To understand how $\overline{C_2}(\infty)$ at $\tilde{B}_{z}/J_{0} = 1$ scales with $N$, we note that there are only $N+1$ orthogonal quantum states for the collective spin $\bm{\Sigma}$, but the Bloch sphere has a surface area of $4\pi$. This is because each orthogonal quantum state occupies a small area on the Bloch sphere with radius $\sim1/\sqrt{N}$ due to the usual uncertainty relation between the different projections of spin. As a result, the upper limit of the integral over $\theta$ in Eq.\,(\ref{eq:C1}) can only reach $\frac{\pi}{2}-\epsilon$, with $\epsilon\sim1/\sqrt{N}$. It can therefore be shown that 
\begin{equation}
\lim_{\frac{\tilde{B}_{z}}{J_0}\rightarrow 1}\overline{C_2}(\infty) \sim \frac{1}{\log(N)}.
\end{equation}
We conclude that size of the dip in the DPT only drops logarithmically with $N$, which may qualitatively explain why only a weak sharpening of the DPT is observed in the experiment as the spin chain grows in size.

\renewcommand{\thefigure}{6}
\begin{figure}
\centering
\includegraphics[width=\columnwidth]{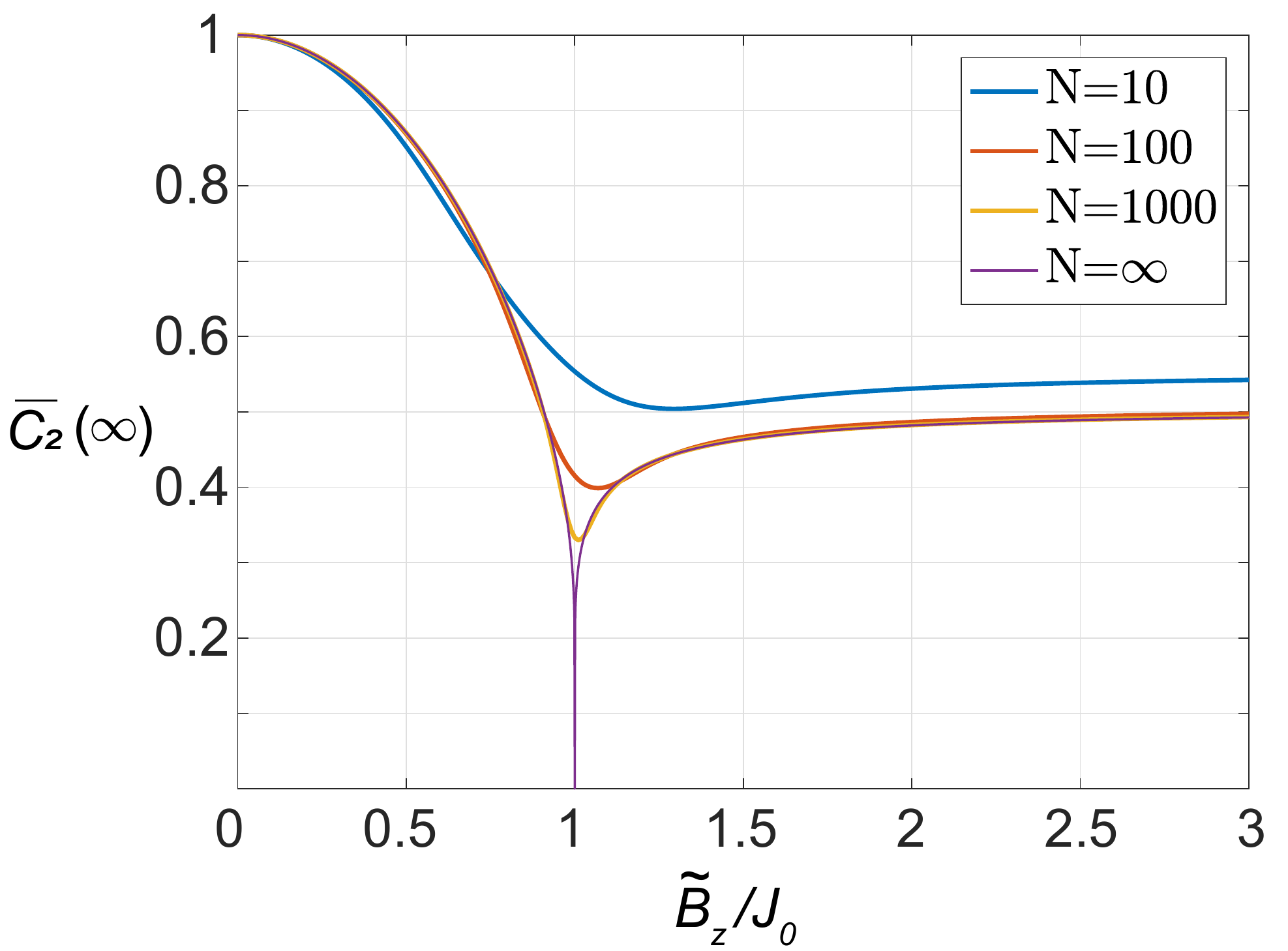}
\caption{The spatially and long-time averaged correlation 
$\overline{C_2}(\infty)\equiv\lim_{T\rightarrow\infty}\frac{1}{T}\int_{0}^{T}\langle(\Sigma^{x}(t)/N)^{2}\rangle dt$
calculated as a function of the ratio $\tilde{B}_{z}/J_{0}$ for the case of $\alpha=0$. The finite $N$ curves are calculated using exact diagonalization, and the $N=\infty$ curve is calculated analytically from Eq. (\ref{eq:C1}).}
\label{fig2}
\end{figure}

\subsection{Acknowledgements}
We acknowledge illuminating discussions with Marko Cetina, Luming Duan, Markus Heyl, Mohammad Maghrebi, Paraj Titum, and Joseph Iosue. This work is supported by the ARO and AFOSR Atomic and Molecular Physics Programs, the AFOSR MURI on Quantum Measurement and Verification, the IARPA LogiQ program, the ARO MURI on Modular Quantum Systems, the ARL Center for Distributed Quantum Information, the NSF Quantum Information Science program, and the NSF Physics Frontier Center at JQI.  G.P. is supported by the IC Postdoctoral Research Fellowship Program.

\bibliographystyle{prsty}

\end{document}